\documentclass[lettersize,journal]{IEEEtran}
\usepackage{amsmath,amsfonts}
\usepackage{algorithmic}
\usepackage{algorithm}
\usepackage{array}
\usepackage[caption=false,font=normalsize,labelfont=sf,textfont=sf]{subfig}
\usepackage{textcomp}
\usepackage{stfloats}
\usepackage{url}
\usepackage{verbatim}
\usepackage{bm}
\usepackage{booktabs}
\usepackage{graphicx}
\usepackage{cite}
\usepackage{colortbl}
\usepackage{xcolor}
\usepackage{tabularx}
\usepackage{tikz}
\usepackage{textcomp}
\usepackage{hyperref}
\usepackage{lipsum}

\newcommand\preprinttext{%
  \footnotesize 
  This work has been submitted to the IEEE for possible publication.
  Copyright may be transferred without notice, after which this version may no longer be accessible.
}
\newcommand\preprintnotice{%
\begin{tikzpicture}[remember picture,overlay]
\node[anchor=south,yshift=5pt] at (current page.south) {\fbox{\parbox{\dimexpr\textwidth-\fboxsep-\fboxrule\relax}{\preprinttext}}};
\end{tikzpicture}%
}
\usepackage{everypage}
\AddEverypageHook{\preprintnotice}

\newcommand{\fmod}{\operatorname{fmod}\left(t, t^{per} \right)}

\begin{document}

\title{Exploring Remote Collaborative Tasks: The Impact of Avatar Representation on Dyadic Haptic Interactions in Shared Virtual Environments}

\author{Genki Sasaki,~\IEEEmembership{Student member,~IEEE,} Hiroshi Igarashi,~\IEEEmembership{Member,~IEEE,}
\thanks{This paper was produced by the IEEE Publication Technology Group. They are in Piscataway, NJ.}
\thanks{Manuscript received April 19, 2021; revised August 16, 2021.}}

\markboth{Journal of \LaTeX\ Class Files,~Vol.~14, No.~8, August~2021}%
{Shell \MakeLowercase{\textit{et al.}}: A Sample Article Using IEEEtran.cls for IEEE Journals}


\maketitle

\begin{abstract}
        This study is the first to explore the interplay between haptic interaction and avatar representation in Shared Virtual Environments (SVEs).
        Specifically, how these factors shape users' sense of social presence during dyadic collaborations, while assessing potential effects on task performance.
        In a series of experiments, participants performed the collaborative task with haptic interaction under four avatar representation conditions: avatars of both participant and partner were displayed, only the participant's avatar was displayed, only the partner's avatar was displayed, and no avatars were displayed.
        The study finds that avatar representation, especially of the partner, significantly enhances the perception of social presence, which haptic interaction alone does not fully achieve.
        However, neither the presence nor the type of avatar representation impacts the task performance or participants' force effort of the task, suggesting that haptic interaction provides sufficient interaction cues for the execution of the task.
        These results underscore the significance of integrating both visual and haptic modalities to optimize remote collaboration experiences in virtual environments, ensuring effective communication and a strong sense of social presence.
\end{abstract}

\begin{IEEEkeywords}
HCI, Avatar Representation, Haptic Interaction, Haptic communication, Remote Collaboration, Virtual human, Social interaction, Collaborative virtual environments, Shared virtual environments
\end{IEEEkeywords}

\section{Introduction}
\IEEEPARstart{V}{i}rtual Environment (VE) technologies provide unprecedented experiences to users.
In particular, Virtual Reality (VR) devices, represented by recent advances in head-mounted displays (HMDs), have become performant and affordable, making VR increasingly accessible to the general public.
In VEs, users can represent their body movements through their avatars\cite{Davis2009-ui}.
Various avatar representations have been reported to alter user experiences\cite{Mohler2010-im}.
Accurate measurement of self-body and reflection in an avatar can enhance the sense of ownership of the body\cite{Waltemate2018-kv}.
Furthermore, accurate body tracking enhances the user experience in VEs\cite{Toothman2019-lr}.

Shared virtual environments (SVEs), where VE is shared with other users, allow users to share space despite each user being remotely located.
SVEs are attracting attention as a new communication tool \cite{Biocca1995-uo}.
In SVEs, visual interaction, which is crucial for users to communicate with each other\cite{Bente2008-sk}, enhances social presence by providing the feeling of ``being there''\cite{Oh2018-iy}.

Several studies have reported that the absence of an avatar representation can reduce social presence\cite{Aseeri2021-xw,Heidicker2017-wk}, a decrease in social presence can potentially cause reduced social communication, which is a critical issue. 
An avatar is a critical element that provide a visual modality in non-verbal communication.
Moreover, remote collaboration using SVEs is expected\cite{Pan2017-dl}.
Remote collaboration through SVEs allows you to solve tasks collaboratively without location constraints, potentially solving issues, such as labor shortages.

Recently, VR applications that integrate haptic technologies have attracted attention.
Haptic feedback can provide information that cannot be conveyed visually and is an essential modality for humans\cite{Baumgartner2013-ep}.
The haptic modality is crucial when touching objects or moving; however, it builds a beneficial communication channel during interactions with others\cite{Groten2013-gt, Roche2022-id}.
In SVEs, haptic interaction provides various benefits to users.
Therefore, haptic interactions influence physical performance as well as the psychological user experience.
Some studies have found that haptic interactions contribute to users feeling a higher social presence\cite{Sallnas2010-tp,Chellali2011-tf,Beelen2013-be,Fermoselle2020-ta,Giannopoulos2008-bv}.
Therefore, haptic interaction is indispensable for performing physical tasks in remote collaboration.

In the course of our research aimed at enabling physical collaborative tasks through remote means, we developed an SVE.
This work enabled us to determine how essential the effects of avatar representation might be in physical remote collaboration.
Although numerous studies have reported that displaying avatars in SVEs can enhance both user experience and physical performance, we wondered if this impact persists when haptic interaction is present.
In particular, in physical remote collaboration, one’s presence is constantly conveyed to others through the haptic modality, raising the question of whether avatar representation remains influential under such circumstances.
Investigating the influence of avatars is important for designing remote collaborative tasks, as it provides valuable guidance for developers and designers.

Nevertheless, it remains unclear whether the display of avatars is essential for physical remote collaboration.
To our knowledge, no study has systematically examined the effect of avatar representation on user experience in the context of haptic interaction.
Specifically, we inquire: Can haptic interaction alone generate sufficient social presence, or is visual avatar representation also required?
We found no empirical evidence that directly addresses this question.

Therefore, in this study, we systematically compare avatar representation in dyadic physical remote collaboration under four conditions (\textit{i.e.} both the user's and the partner's avatars are displayed, only the user's avatar is displayed, only the partner's avatar is displayed, and no avatars are displayed).
We measured social presence, task performance, and users' force effort.
The results showed that participants felt a decrease in social presence when the partner's avatar was not displayed.
Meanwhile, task performance and effort remained constant under all conditions.
These findings suggest that although haptic interaction alone does not fully compensate for the absence of visual cues in terms of social presence, it does not hinder task completion.
Consequently, participants found the social presence provided by haptic interactions to be insufficient; however, the task execution was not affected by the representation of avatars.

In this study, we used a generic, gender-neutral avatar to remove confounding variables, such as voice or distinctive features, thereby isolating the effect of avatar presence on the interplay between visual and haptic cues.
However, this minimalistic approach may limit the generalizability of our findings to scenarios involving realistic or personalized avatars with additional communication cues (\textit{e.g.}, voice or facial expressions).

Based on these results, we emphasize the necessity of avatar representation for developing physical remote collaboration especially for maintaining social presence.
The contributions of this paper are as follows:
1) If the purpose is solely task execution in physical remote collaboration, avatar representation can be omitted without compromising performance.
2) In social communication contexts, visual interaction through avatar representation is more critical than haptic interaction for maintaining social presence.

\section{RELATED WORK}
\subsection{Avatar representation in shared virtual environment}
In social interactions, visual information plays a crucial role.
Especially in SVEs provided through HMDs, users expect a bodily experience that mimics real-world movements and synchronizes with their own actions.
Avatars, as a form of visual information, offer users a realistic representation of their bodies, as well as facilitate the presence of others\cite{Weidner2023-ee}.
The presence of avatars of others, compared to audio and video communication channels, sufficiently enhances the user's experience of emotional intimacy\cite{Bente2008-sk}.
The representation of avatars plays a crucial role in determining user experience and task performance\cite{Nowak2003-qa}.
Gao \textit{et al.} reported that the use of full-body avatars improved task performance compared to using only hands or hands and torso\cite{Gao2020-kc}.
Moreover, Pan \textit{et al.} conducted competitive and collaborative tasks in VEs with and without self-avatars and in face-to-face settings\cite{Pan2017-dl}.
They found that the absence of a self-avatar during collaborative tasks caused a decrease in task scores.
However, no significant differences were observed in competitive tasks.
Thus, the impact of avatar representation on performance may vary depending on the context of the task.

Furthermore, avatar representation contributes to enhancing social presence.
Aseeri \textit{et al.} compared three conditions: no avatar, scanned avatar, and real avatar, and found that scanned avatar and real avatar demonstrated high levels of social presence\cite{Aseeri2021-xw}.
In addition, complete mapping\cite{Heidicker2017-wk}, full-body tracking\cite{Yoon2019-gi}, and facial expressions\cite{Kang2022-ee} in avatars improve social presence.
In summary, representing avatars in SVE significantly affects users and contributes strongly to the formation of communication channels.

\subsection{Social presence and haptic interaction in shared virtual environment}
Haptic interaction allows users to share forces among themselves, enabling them to perform physical remote collaboration.
Physically connected dyads form communication channels through the exchange of forces, allowing them to receive information from their partner \cite{Groten2013-gt}.
Such interactions offer various physical benefits to dyads: improvement in time to motor learning \cite{Ganesh2014-gk, Kim2022-dr}, enhancement of performance \cite{Reed2008-gc}, maintenance of balance \cite{Wu2021-jz}, and negotiation of strategies \cite{Groten2013-gt}.

They also allow user to experience benefits beyond physical advantages.
Several studies have investigated the impact of haptic interactions in SVEs by toggling the presence and absence of such interactions among dyads \cite{Sallnas2010-tp, Chellali2011-tf, Beelen2013-be, Fermoselle2020-ta, Giannopoulos2008-bv}.
Chellali \textit{et al.} conducted a task in which dyads trained in needle insertion procedures using a Follow--Follower system, and demonstrated that haptic interactions improved performance and co-presence\cite{Chellali2011-tf}.
Additionally, Sallnas \textit{et al.} found that exchanging objects through haptic interactions enhanced social presence\cite{Sallnas2010-tp}.

Therefore, haptic interactions provide both physical and psychological benefits to users, including improved motor learning, better performance, and enhanced social presence.
In particular, the social presence fostered by haptic interactions supports the idea that they can serve as a communication channel in social interaction contexts.
Meanwhile, previous findings indicate that visual interaction through avatar representation enhances social presence and task performance. 

Focusing on physical remote collaboration using an SVE, haptic communication is indispensable and realized through haptic feedback, whereas avatar representation likely serves as the visual communication channel.
Additionally, verbal communication may be involved.
Thus, in SVE-based physical remote collaboration, social interaction is supported concurrently by multiple channels.
Consequently, a critical question arises: do performance and social presence depend on a single channel, all channels, or specific combinations of channels?
Furthermore, if one channel is lost, do performance and social presence decline, or are they preserved through the others?
To the best of our knowledge, no research has specifically examined the impact of avatar representation in the context of physical remote collaboration.
Our study is the first to address this question.

\section{METHODS}
\subsection{Research hypothesis}
This study investigates the impact of avatar representation in remote physical collaboration.
In particular, previous studies have highlighted the roles of haptic interaction and task context in shaping user experience, social presence, and performance in SVEs or remote settings. 
Therefore, we propose the following hypotheses to elucidate how different forms of avatar representation might affect various aspects of collaborative work—even when participants are physically apart.

\begin{itemize}
        \item H1: Previous studies in SVE have reported that social presence can be heightened by perceiving the existence of others through haptic modalities\,\cite{Sallnas2010-tp, Chellali2011-tf, Beelen2013-be, Fermoselle2020-ta, Giannopoulos2008-bv}. This may be equivalent to the social presence arising from linguistic communication or visual interaction (\textit{e.g.}, avatars)\,\cite{Aseeri2021-xw} that have been noted in prior work. In physical remote collaboration using SVE, even if an avatar is not displayed, users can continue to feel their partner’s presence through haptic interaction. Therefore, regardless of the avatar representation, users can maintain the same level of social presence.
        
        \item H2: It has been reported that the effect of avatar representation on users differs depending on the task context in which they engage within SVE\,\cite{Pan2017-dl}. In a context that requires intensive focus on task execution, the user is primarily devoted to achieving performance, thereby diminishing the influence of avatar representation. In physical remote collaboration, where the goal is to accomplish a particular task, the dyad may continue to achieve their highest possible performance regardless of avatar representation. Consequently, performance remains constant irrespective of the avatar representation.
        
        \item H3: In physical collaborative tasks, it has been reported that force is used to perform the task as well as to negotiate intentions with one’s partner\,\cite{Groten2013-gt}. Considering this social role of force, the psychological effects induced by changes in avatar representation may ultimately alter the way force is applied.
\end{itemize}

\subsection{Experimental design}
\begin{figure*}[t]
        \begin{center}
                \includegraphics[width=180mm]{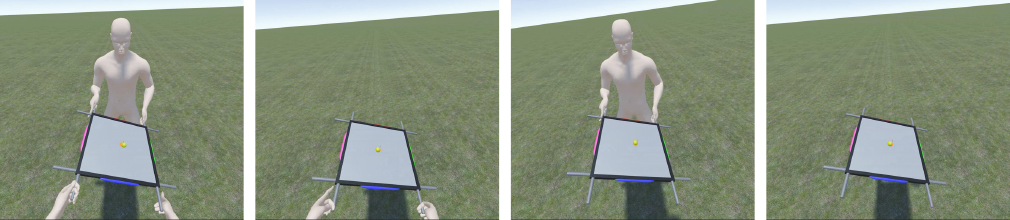}
                \caption{Participant's view on each condition of avatar representation. From left to right, are ALL, ALONE, OTHER, and NONE.}
                \label{fig:condition}
        \end{center}
\end{figure*}

To assess the impact of avatar representation, we compared four conditions of avatar representation:
\begin{itemize}
        \item ALL: Both participant's and partner's avatars were displayed in the VE. It assumed that both users had body tracking devices.
        \item ALONE: Only participant's avatar was displayed. Although both users had body tracking devices, avatar-related data was not transmitted to the partner. This condition could save communication bandwidth, ensuring reliable transmission of haptic information.
        \item OTHER: Only the partner's avatar was displayed.
        \item NONE: No avatars were displayed. This simplifies and reduces the cost of SVE development as no body-tracking device is required.
\end{itemize}

The conditions ALL, ALONE, and NONE are those anticipated in real environments.
However, OTHER has been included based on our previous research \cite{Sasaki2022-pf}, but it is a condition that does not occur in reality (as only the partner receives body tracking information, and no avatar is displayed on one's own screen).
We were interested in the effects that occur when self-avatar is not displayed.
If self-avatar alone is not displayed in the SVE, can participants maintain their sense of social presence?
In addition, each condition was applied simultaneously to both participants.

\subsection{Apparatus and Implementation}
\begin{figure*}[t]
        \begin{center}
                \includegraphics[width=180mm]{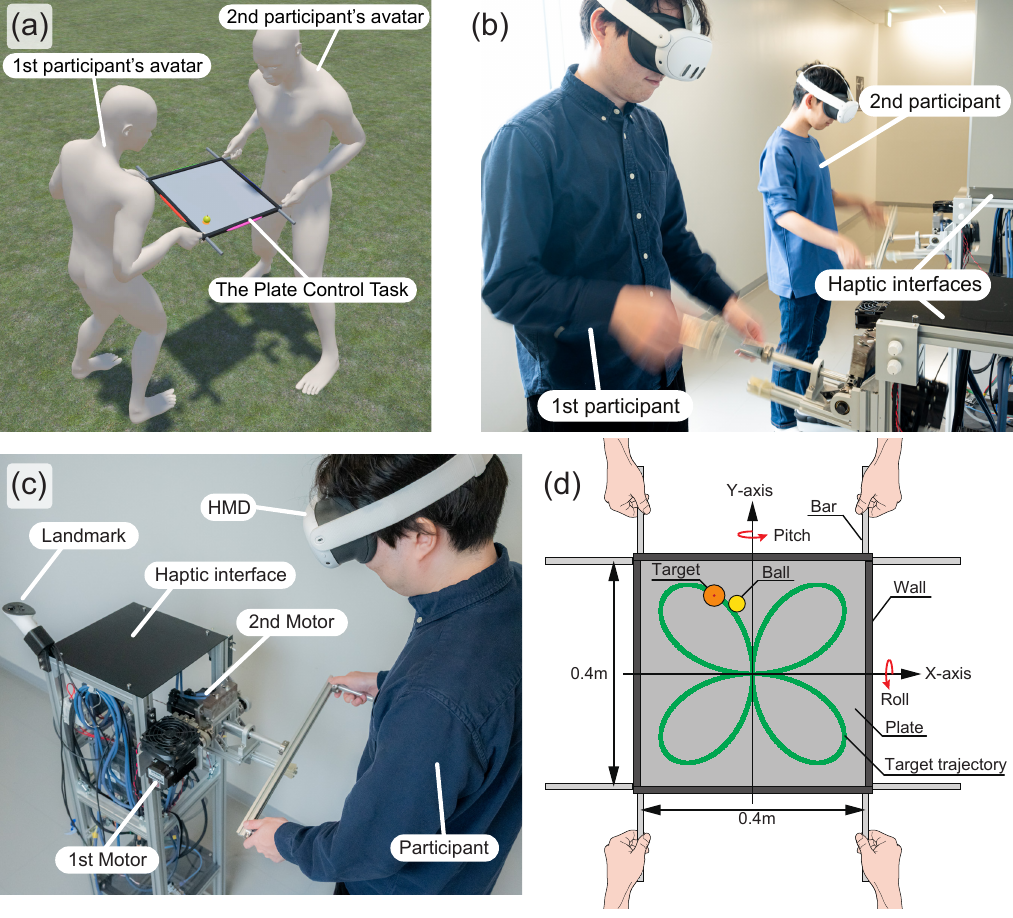}
                \caption{Experimental setup. (a) The shared virtual environment. (b) Participants could perform the collaborative task through the independent haptic interface. (c) Participant install the HMD and see the SVE. (d) The plate control task.}
                \label{fig:setup}
        \end{center}
\end{figure*}

We developed an experimental environment for a physical remote collaboration task as shown in Fig. \ref{fig:setup}.
The SVE displayed a plate and avatars representing the participants' body movements.
The dyads manipulated the plate through a haptic device, performing collaborative tasks.

\subsubsection{Shared virtual environments}
Our developed SVE was created using Unity 2022.3.23f1.
In this SVE, a plate was fixed in a floating position at a height of 1 m, surrounded by avatars.
Dyad's avatars were positioned face-to-face across the plate.
The plate is a square with each side 0.4 m.
Participants accessed the SVE using the Meta Quest 3, which tracked the position of participants' heads, wrists, and each joint of the fingers, applying these positions in real time to the avatars' actions.
Data from Meta Quest 3 were captured using the Meta XR Core SDK v63.0.0.
The movements of the avatar's body parts not derived from captured positions were determined using Final IK.
Inverse Kinematics (IK) techniques can cause errors, ultimately causing a feeling of discomfort for participants\cite{Yun2023-ax}.
However, because the participants moved within a limited range in this study, these errors were not sufficiently significant to affect the experiment; thus, they were not be considered.

The employed avatars were generic with no gender or distinctive features, reflecting only participants' height as physical information. 
Their overall skin color was uniform, exhibiting a cream-like hue that falls between yellow and white, and textures were smooth with no variation across body parts. 
In addition, there were no prominent features in the chest or groin area, making it difficult to visually identify a particular gender. 
Thus, the influence of sex or facial personalization was minimized. 
However, the force exerted on an avatar has been reported to decrease when it appears female \cite{Bailenson2008-kn}.
Furthermore, participants interacting with expressive avatars perceive stronger social presence and attractiveness, exhibiting better task performance than those with non-expressive avatars \cite{Wu2021-pq}. 
We adopted this minimalistic design to isolate fundamental effects of avatar representation and haptic interaction, avoiding confounds introduced by realistic or expressive avatars.

\subsubsection{Haptic Interface}
Haptic interactions between dyads were bilateral.
One of these haptic interfaces is provided for each participant and operates independently.
The haptic interface featured two motors, and had a gimbal mechanism that could rotate along the roll and pitch axes.
Thus, participants could manage two degrees of freedom of movement.
The motors used were Mitsubishi Electric's HG-MR43, and to maintain high backdrivability, they were driven directly.
Direct drive is often adopted in the research of physical Human--Human Interaction (pHHI) \cite{Melendez-Calderon2011-wz, Roche2022-id}.
Extending from the gimbal mechanism was a handle that could be gripped with both hands.
The length of the handle from the left to the right was 0.4 m while that to the gimbal mechanism was 0.2 m, perfectly corresponding to the length of the plate.
The gimbal mechanism was fixed at a height of 1 m and the participants operated it through the haptic interface in a standing position.

\subsubsection{Haptic interaction between dyad}
To provide haptic interaction to the dyad, we applied acceleration-based admittance control to the haptic interfaces, allowing bilateral driving \cite{Nagatsu2024-ku}. 
This form of control, suitable for establishing haptic interactions, enables precise force transmission between participants \cite{Roche2018-gg}. 
Bilateral control represents motion transfer by ensuring that actuators satisfy Newton's third law of action and reaction, as well as positional consistency \cite{Yokokohji1994-hd}. 
Owing to these motion transfer properties, bilateral control has been widely adopted in various teleoperation systems \cite{Iida2004-on,Hirche2007-jg,Salvietti2017-iu}. 
To achieve both force and positional transmission with a single actuator, force-based and position-based motions must be integrated at the acceleration level \cite{Hannaford1989-hg}. 
In this control scheme, the desired acceleration is provided to the control system as the command, and the system’s response acceleration must match this target acceleration. 
We employed a disturbance observer to implement acceleration-based control \cite{Ohnishi1996-ii}, which suppresses external forces in the control system, aligning the command acceleration with the controlled acceleration. 
In addition, a reaction torque observer was used to measure participant forces without a force sensor \cite{Murakami1993-pd}.

Admittance control defines how a system responds to human operator forces through its dynamics, thereby allowing designers to freely specify these dynamics. 
This capability allows for it to provide haptic interactions as though the dyads are jointly manipulating a single object in the real world, a method used in pHHI studies \cite{Groten2013-gt}. 
In this study, the dynamics of the plate were shared between the two participants.

Here, the forces that each dyad member applied through their respective haptic interfaces generated the actual force acting on the plate:
\begin{equation}
        {\bm \tau^{plt}} = 
        \left[\begin{matrix} 
                \tau_1^{p} + \tau_2^{p} \\
                \tau_1^{r} + \tau_2^{r}  
        \end{matrix} \right]
        ,
        \label{eq:tau_plt}
\end{equation}
where $\tau_1$ and $\tau_2$ denote each participant's force input, with the superscripts $p$ and $r$ indicating the pitch and roll axes, respectively.

The plate’s dynamics are expressed as:
\begin{equation}
    {\bm \tau^{plt}} = M \ddot{\bm{\theta}}^{plt} + D \dot{\bm{\theta}}^{plt}
    ,
\end{equation}
where $\ddot{\bm \theta}^{plt}$ and $\dot{\bm{\theta}}^{plt}$ represent the plate's angular acceleration and velocity vectors, respectively, which are integrated to obtain the plate's angular position vector $\bm{\theta}^{plt}$.
These vectors are defined as follows:
\begin{align}
        \ddot{\bm \theta}^{plt} &= \left[\begin{matrix} \ddot{\theta}^{plt} & \ddot{\phi}^{plt} \end{matrix} \right]^T \notag \\
        \dot{\bm \theta}^{plt} &= \left[\begin{matrix} \ddot{\theta}^{plt} & \ddot{\phi}^{plt} \end{matrix} \right]^T \\
        {\bm \theta}^{plt} &= \left[\begin{matrix} \theta^{plt} & \phi^{plt} \end{matrix} \right]^T \notag
        .
\end{align}
where $\theta$ and $\phi$ are the angles around the pitch and roll axes, respectively. 
In the experiment, the mass $M$ and the viscous coefficient $D$ of the plate were set to 3.28 and 1.5, respectively.

Through admittance control, participants can jointly manipulate a single shared object, determining its motion by the forces they apply—just as objects behave in the real world. 
If no haptic feedback is provided (\textit{i.e.}, no haptic interaction), the plate’s position reflects the average of each participant’s manual inputs, causing a discrepancy between the visually perceived plate position and participants’ actual movements. 
Such conditions are not appropriate for executing physical collaborative tasks. 
Therefore, in this study, we consistently provided haptic feedback to ensure realistic interaction.

\subsubsection{Plate control task}
A ball and a target were placed on the plate.
The dyads were instructed to position the ball on the target.
This task, which we call the plate control task, required precise control and coordination.

The trajectory of the target followed a rose curve, which moved according to the progression of time.
The movement of the target was calculated as:
\begin{equation}
        {\bm x^{tgt}} = r
        \left[
        \begin{matrix}
                \sin\left(4\pi \zeta \right) 
                \cos\left(2\pi \zeta \right) \\
                \sin\left(4\pi \zeta \right) 
                \sin\left(2\pi \zeta \right)
        \end{matrix}
        \right]
        .
        \label{eq:rose curve}
\end{equation}
Here, ${\bm x^{tgt}} = \left[ \begin{matrix} x^{tgt} & y^{tgt} \end{matrix} \right]^T$ represents the position vector of the target on the plate, indicating its x and y axes position.
$\zeta = \fmod / t^{per}$ is the quotient remainder of $t$ and $t^{per}$ divided by $t^{per}$, which is a ratio of 0.0 to 1.0.
This can be computed by using the fmod function in C, Pyhon, etc.
$t$ represents time, and $t^{per}$ is the period of the rose curve from start to finish, designed to be 20 s in this study.
$r$ is the radius of the trajectory, set to 0.2 m.
Dyads were not shown the trajectory visually during the experiment but memorized it during practice sessions.

The ball moved according to the plate's inclination, governed by the dynamics applied as:
\begin{equation}
        {\bm x^{bll}} = mg\sin{\bm \theta^{plt}} - \eta \dot{\bm x}^{bll}
        .
        \label{eq:ball}
\end{equation}
Here, ${\bm x^{bll}} = \left[ \begin{matrix} x^{bll} & y^{bll} \end{matrix} \right]^T$ is the position vector of the ball on the plate. $\dot{\bm x}^{bll}$ is the velocity vector of the ball. 
$m$ is the ball mass, $g$ is the gravitational acceleration, $\eta$ is the viscosity coefficient, with values set to 0.014, 9.8, and 1.5, respectively, in this study. 
$m$ and $\eta$ are critical parameters determining the movement of the ball relative to the tilt of the plate, thereby significantly influencing the difficulty of the task.
These parameters were chosen experimentally to set the difficulty of the task.

\subsubsection{Co-localization}
Synchronizing the positions of the haptic interface and the VE has significant implications on task performance and user perception, as noted in previous studies \cite{Swapp2006-qm, Saint-Aubert2020-lz}.
Ensuring that the haptic interface and the plate displayed within the SVE were precisely aligned was crucial.
To facilitate this, a Meta Quest 3 controller was attached to the top of the haptic interface, serving as a landmark to provide exact positional data.

Before starting the experiments, participants were calibrated to align the landmark with the plate's position in the SVE.
This setup allowed participants to manipulate the SVE plate through the haptic interface without any sense of discrepancy or discomfort.

The landmark was used to share the positions of avatars.
When a participant grasped the plate, it was essential that the avatar of their partner appeared to be grasping the plate as well, thereby enhancing the sense of collaboration and presence within the SVE.
The avatar's position was captured from the landmark's local coordinate system and transmitted to other devices, ensuring that the dyad could share the space around the plate completely, irrespective of the interface's physical location.
The avatar position data were updated and transmitted every 10 milliseconds, allowing participants to focus on their tasks without experiencing any perceivable delay.
This technology facilitated a seamless integration of physical and virtual interactions, crucial for maintaining high levels of engagement and performance in collaborative virtual tasks.

\subsection{Participants}
We performed a priori power analysis using G*Power \cite{Faul2009-ul} to determine our sample size. 
To detect a medium effect size, we selected an effect size of \textit{f} = 0.25, an alpha error probability of 0.05, and a power of 0.8.
Because we planned to conduct a repeated measures ANOVA with four experimental conditions, we set the number of groups to one while of measurements to four. 
Consequently, 28 participants were included in this experiment, all university students.
Participants provided informed consent according to the Declaration of Helsinki before the experiment.
The experiment was approved by the Human Life Ethics Committee of Tokyo Denki University.
Among the participants, 24 were right-handed and four left-handed.
The group included 24 males and 4 females aged 21 to 28 ($M=22.96$, $SD=2.08$).

Participants were randomly paired into 14 dyads, with relationships ranging from friends and acquaintances to strangers.
The visibility of a friend's face during a collaborative task has been reported to influence performance \cite{Boyle1994-tu}.
However, the absence of visibility of the face does not affect the results, even with close friends \cite{Brennan2015-sl}.
Therefore, the relationship between partners was not considered in this experiment.
All participants were first briefed on the experiment's objectives, followed by an explanation of the equipment used.

\subsection{Procedure}
Initially, participants provided personal information (biological sex, age, height, and dominant hand) to collect physical data.
The participants practiced alone to familiarize themselves with the plate control task, SVE, and HMD.
They wore the Meta Quest 3 and calibrated it to synchronize the haptic interface with the VE, verifying that their avatar tracked their body, hands, and fingers accurately.
During practice, participants could ask the experimenter questions about the plate control task without restrictions.
The practice continued until the participants declared that they were comfortable with the plate control task, with no time limits set by the experimenter.

After practice, each dyad collaboratively performed the plate control task for 80 s.
The avatar representation condition was applied randomly from four conditions, and visual interaction between the dyads was presented.
During the experiment, verbal communication between the dyads was prohibited by the experimenter.
Following the plate control task, participants removed the Meta Quest 3 and completed a questionnaire as instructed by the experimenter (details in \ref{sec:Survey}).
The questionnaire was displayed on a computer.
Because we employed a within-subjects design, all participants were required to experience each of the four conditions.

After completing the questionnaire, the participants were provided with at least a two-hour rest to account for fatigue and memory of their responses. 
On each day of participation, they completed two of the four experimental conditions, thus, they requiring two days to finish all conditions.
Consequently, each participant spent approximately four hours per day in the experiment, that is, eight hours over the two days. 
No restrictions were placed on participants' activities during the test. 
After the rest, participants performed the plate control task under a different condition and completed the questionnaire again. 
The experimenter conducted the plate control task and questionnaire four times with each pair to compare the four conditions.

\subsection{Measures}
A systematic review was conducted on the effects of avatars and agents during the use of HMD highlights that avatars can be effective or ineffective, depending on the task and situation \cite{Weidner2023-ee}.
Furthermore, they also point out that combining subjective evaluations with numerical data enhances the validity of research findings.
Bailenson \textit{et al.} argue that subjective measures, self-reported, depend on the participant's ability to accurately assess their experiences \cite{Bailenson2004-ei}.
Subjective measures are susceptible to participant bias, which can affect reliability. However, objective measures can detect responses to digital representations, such as personal distance, which are not captured by self-reports, suggesting higher sensitivity.
Thus, in this study, user experience (social presence) is subjectively measured, while task-related metrics (performance and effort) are objectively measured.

\subsubsection{Subjective measures}\label{sec:Survey}
Social presence is the psychological sense that another person is ``being there'' when using media or communication technologies \cite{Biocca1997-fo}.
This concept delivers the intimacy and immediacy of real interactions despite physical separation.
A higher richness of information in communication media can create a stronger sense of social presence.
To deepen understanding of how people feel connected in VEs, social presence is often used \cite{Oh2018-iy}.

To measure social presence, this study used the Networked Minds Social Presence Questionnaire (NMSPQ) \cite{Biocca2001-jv, Biocca2003-dj}.
The NMSPQ includes five sub-scales: Co-presence, Perceived Attentional Engagement, Perceived Emotional Contagion, Perceived Comprehension, and Perceived Behavioral Interdependence.
Because the interaction in our task was limited to non-verbal communication, participants had no opportunity for conversation and could not comprehend their partner's thoughts.
In addition, the general avatars used were expressionless and lacked personality, which prevented the transmission of emotions to the partner.
Therefore, Perceived Comprehension and Perceived Emotional Contagion were removed from the questionnaire.

Consequently, only the relevant sub-scales of Co-presence, Perceived Attentional Engagement, and Perceived Behavioral Interdependence were retained.
The questions were adjusted to align with the experimental context, and the questionnaire comprised 20 items.
Responses were collected on a 7-point Likert scale ranging from ``Strongly disagree'' to ``Strongly agree.''

\subsubsection{Objective measures}
In this study, performance and effort were used as objective measures.
The objective of the plate control task was to move a ball on a target, which required the dyads to collaborate to decrease the distance between the target and the ball.
We decided to measure performance by calculating the root-mean-squared error (RMSE) between the target and the ball.
Performance is defined as follows:
\begin{equation}
        P = \frac{\sum_{k=1}^{N} \left|\left|{\bm x_k^{tgt}} - {\bm x_k^{bll}}\right|\right|}{\sqrt{N}}
        .
        \label{eq:performance}
\end{equation}
Here, $k$ represents the number of steps ($k=1\ldots N$), corresponding to the number of data points captured from the haptic interface.
Note that a lower value of $P$ indicates better performance.

Effort was measured.
Although participants applied force to move the plate, not all applied force effectively contributed to the movement.
According to Newton's laws, when two equal forces are applied in opposite directions to an object, they cancel out, and the object does not move.
However, if forces are applied in the same direction, the object moves.
Thus, the difference between the total forces applied to the object and the forces that actually moved the object represents the effort not contributing to the movement, considered as effort here.
Although effort does not directly impact performance, it is considered crucial for haptic communication used by dyads to negotiate for achieving high performance.

The total force applied to the plate is denoted by:
\begin{equation}
        {\bm \tau^{tot}} = 
        \left[\begin{matrix} 
                \left|\tau_1^{p}\right| + \left|\tau_2^{p}\right| \\
                \left|\tau_1^{r}\right| + \left|\tau_2^{r}\right|
        \end{matrix} \right]
        .
        \label{eq:tau_tot}
\end{equation}
From which the definition of effort follows:
\begin{equation}
        F^{eff} = \frac{1}{N \cdot L} \sum_{k=1}^{N}
        \left[
        \begin{matrix}
                1 & 1
        \end{matrix}
        \right] 
        \left({\bm \tau_k^{tot}} - \left|{\bm \tau_k^{plt}}\right|\right)
\end{equation}
Here, $k$ represents the number of steps ($k=1\ldots N$), and $L$ is the half length of the plate used to convert torque into force.
This measure helps to understand the extent of unproductive effort exerted by participants during the task.

\section{RESULTS}
Although participants performed the tasks in dyads, we treated each individual's data as an independent sample for our analyses. 
This study aimed to compare individual evaluations and responses, rather than to analyze interactions within each dyad.
Therefore, we did not employ a nested (multi-level) analysis that accounts for dyadic structure and instead performed a standard ANOVA to examine main effects. 
However, owing to the nature of the objective measures, we could not separate the data by individual; hence, we used one dataset per dyad for those measures.

Because participants could potentially observe each other's faces and be influenced by sound or smell, there was a possibility of mutual interference. 
Nevertheless, our tasks and metrics were designed to be evaluated on an individual basis, and we did not anticipate systematic interactions sufficiently strong to bias the results (\textit{e.g.}, one participant’s responses heavily shaping the other’s). 
A preliminary examination of the data revealed no notable correlations or clustering among participants, reinforcing our decision not to model the dyad factor.

In the box plots presented in this study, the central line indicates the median, and the box edges represent the first and third quartiles. 
The whiskers show the range excluding outliers, determined using 1.5 times the interquartile range (IQR).

\subsection{Subjective measures}
Average scores were calculated for the three selected NMSPQ subscales, as well as for overall Social Presence (defined as the mean of all subscale scores).
To verify whether Social Presence, Co-presence, Perceived Attentional Engagement, and Perceived Behavioral Interdependence followed normal distributions, the Shapiro–Wilk test was performed.
No significant deviations from normality were found for any of the scales ($p > .05$), indicating that these data were suitable for parametric analysis.

\begin{figure*}[t]
        \begin{center}
                \includegraphics[width=\linewidth]{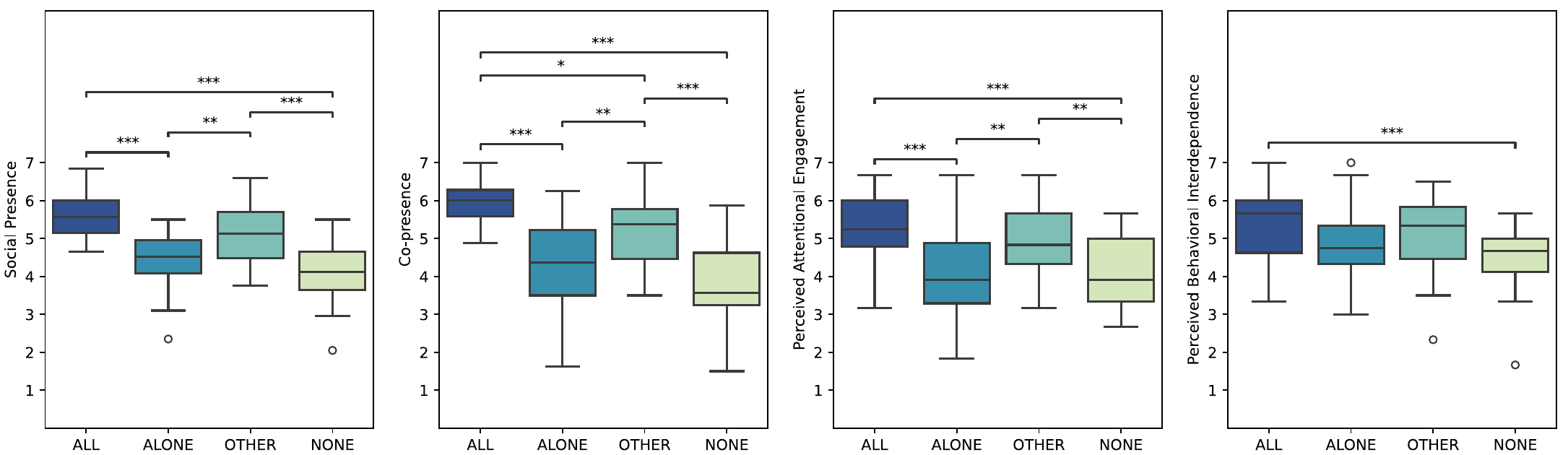}
                \caption{
                        Subjective measures of Social Presence (NMSPQ). From left to right, Social Presence, Co-presence, Perceived Attentional Engagement and Perceived Behavioral Interdependence (*: \textit{p} $<$ 0.05, **: \textit{p} $<$ 0.01, ***: \textit{p} $<$ 0.001).
                        }
                \label{fig:survey}
        \end{center}
\end{figure*}

Subsequently, repeated measures ANOVA was conducted to compare scores under the four conditions (ALL, ALONE, OTHER, and NONE) for each scale.
Mauchly’s test of sphericity indicated violations for Social Presence (\textit{p} = .009) and Co-presence (\textit{p} = .002), but not for Perceived Attentional Engagement (\textit{p} = .994) or Perceived Behavioral Interdependence (\textit{p} = .085).
Consequently, the Greenhouse–Geisser correction was applied for Social Presence and Co-presence.
The repeated measures ANOVA revealed significant main effect in Social Presence (\textit{F}(2.296, 61.985) = 24.442, \textit{p} $<$ .001, $\eta^2_p$ = .475), Co-presence (\textit{F}(2.069, 55.858) = 27.301, \textit{p} $<$ .001, $\eta^2_p$ = .503), Perceived Attentional Engagement (\textit{F}(3, 81) = 13.404, \textit{p} $<$ .001, $\eta^2_p$ = .332), and Perceived Behavioral Interdependence (\textit{F}(3, 81) = 5.807, \textit{p} = .001, $\eta^2_p$ = .177).
To compare all group pairs, post hoc tests with Bonferroni-corrected \textit{p}-values were performed.
Detailed post hoc results are presented in Table \ref{tab:postHocComparisons-SocialPresence} and Fig. \ref{fig:survey}.

\begin{table}[t]
        \footnotesize
	\centering
	\caption{Post Hoc Comparisons - Social Presence}
	\label{tab:postHocComparisons-SocialPresence}
        \begin{tabular}{
                ll|
                >{\raggedleft\arraybackslash}p{0.8cm}
                >{\raggedleft\arraybackslash}p{0.5cm}
                >{\raggedleft\arraybackslash}p{0.8cm}
                >{\raggedleft\arraybackslash}p{0.8cm}
                >{\raggedleft\arraybackslash}p{0.9cm}
                }
                \toprule
                \multicolumn{2}{l}{Pair} & \multicolumn{1}{c}{\textit{MD}} & \multicolumn{1}{c}{\textit{SE}} & \multicolumn{1}{c}{\textit{t}} & \multicolumn{1}{c}{\textit{d}} & \multicolumn{1}{c}{\textit{p}$_{bonf}$} \\
                \rowcolor{lightgray} \multicolumn{7}{c}{Social Presence} \rule[0pt]{0pt}{8pt} \\
                ALL & ALONE & 1.148 & .186 & 6.158 & 1.521 & $<$ .001\\
                & OTHER & 0.489 & .186 & 2.624 & 0.648 & .062\\
                & NONE & 1.448 & .186 & 7.764 & 1.918 & $<$ .001\\
                ALONE & OTHER & -0.659 & .186 & -3.534 & -0.873 & .004\\
                & NONE & 0.300 & .186 & 1.609 & 0.397 & .669 \\
                OTHER & NONE & 0.959 & .186  & 5.143 & 1.270 & $<$ .001 \\
                \rowcolor{lightgray} \multicolumn{7}{c}{Co-presence} \rule[0pt]{0pt}{8pt}\\
                ALL & ALONE & 1.612 & .257 & 6.281 & 1.604 & $<$ .001\\
                & OTHER & 0.728 & .257 & 2.836 & 0.724 & .035\\
                & NONE & 2.143 & .257 & 8.351 & 2.133 & $<$ .001\\
                ALONE & OTHER & -0.884 & .257 & -3.445 & -0.880 & .005\\
                & NONE & 0.531 & .257 & 2.070 & 0.529 & .250 \\
                OTHER & NONE & 1.415 & .257  & 5.515 & 1.409 & $<$ .001 \\
                \rowcolor{lightgray} \multicolumn{7}{c}{Perceived Attentional Engagement} \rule[0pt]{0pt}{8pt}\\
                ALL & ALONE & 1.179 & .223 & 5.287 & 1.222 & $<$ .001\\
                & OTHER & 0.304 & .223 & 1.362 & 0.315 & 1.000\\
                & NONE & 1.071 & .223 & 4.807 & 1.111 & $<$ .001\\
                ALONE & OTHER & -0.875 & .223 & -3.926 & -0.907 & .001\\
                & NONE & -0.107 & .223 & -0.481 & -0.111 & 1.000 \\
                OTHER & NONE & 0.768 & .223  & 3.445 & 0.796 & $<$ .005 \\
                \rowcolor{lightgray} \multicolumn{7}{c}{Perceived Behavioral Interdependence} \rule[0pt]{0pt}{8pt}\\
                ALL & ALONE & 0.500 & .218 & 2.292 & 0.534 & .147\\
                & OTHER & 0.357 & .218 & 1.637 & 0.381 & .633\\
                & NONE & 0.899 & .218 & 4.120 & 0.959 & $<$ .001\\
                ALONE & OTHER & -0.143 & .218 & -0.655 & -0.152 & 1.000\\
                & NONE & 0.399 & .218 & 1.828 & 0.426 & 0.427 \\
                OTHER & NONE & 0.542 & .218  & 2.483 & 0.578 & 0.091 \\
                \bottomrule
        \end{tabular}

        \vspace{1.0ex}
	\raggedright 
	\footnotesize
        \textit{Note.}
        \textit{MD}, \textit{SE}, \textit{t}, \textit{d} and \textit{p}$_{bonf}$ stand for Mean Difference, Standard Error, \textit{t}-value, Cohen's \textit{d} and \textit{p}-value with Bonferroni correction, respectively.
        \textit{p}-value adjusted for comparing a family of 6.
\end{table}

\subsection{Objective measures}
\begin{figure}[t]
        \begin{center}
                \includegraphics[width=0.8\linewidth]{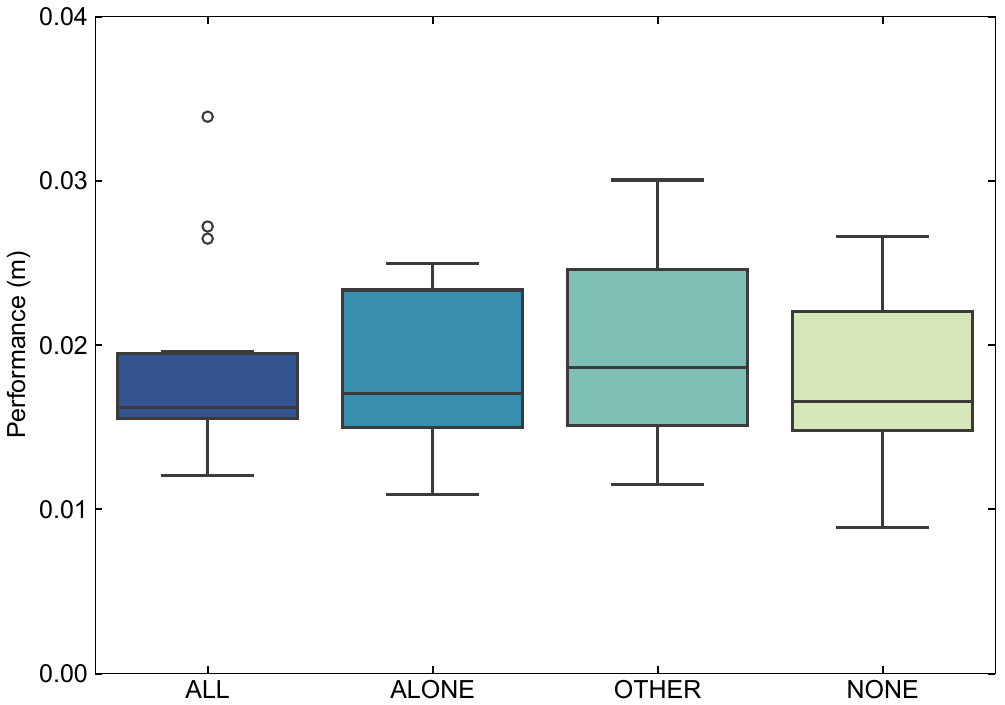}
                \caption{Comparison of task performance of the plate control task.}
                \label{fig:performance}
        \end{center}
\end{figure}

\begin{figure}[t]
        \begin{center}
                \includegraphics[width=0.8\linewidth]{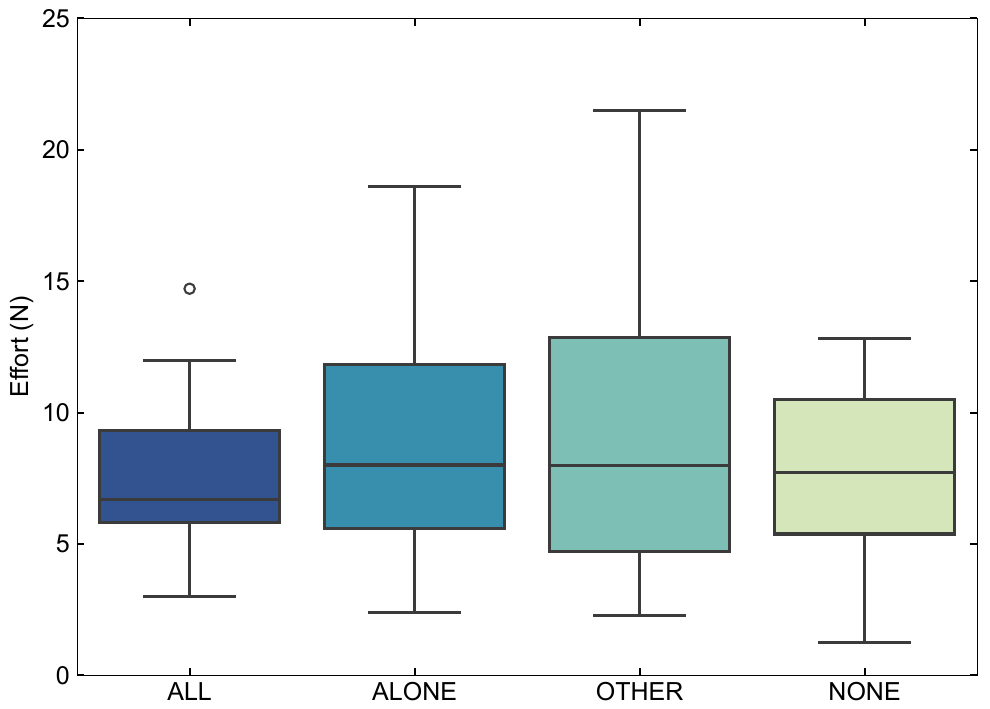}
                \caption{Comparison of effort of the plate control task.}
                \label{fig:ine}
        \end{center}
\end{figure}

For performance and effort, the Shapiro–Wilk test was performed for each condition (ALL, ALONE, OTHER, NONE).
Performance showed a significant deviation from normality (\textit{p} = .009), whereas effort did not (\textit{p} = .796).
Accordingly, the Friedman test (a nonparametric equivalent to repeated measures ANOVA) was used to assess whether performance differed across the four conditions.
No significant main effect were identified ($\chi^2$(3) = 7.629, \textit{p} = .054, \textit{W} = .182).

Because effort met normality assumptions, repeated measures ANOVA was employed to test for differences across conditions.
Mauchly’s test of sphericity for effort was no significant (\textit{p} = .514).
The ANOVA indicated no significant main effect (\textit{F}(3, 39) = 1.375, \textit{p} = .265, $\eta^2_p$ = .096).
Comparisons of performance and effort under the different conditions are illustrated in Fig. \ref{fig:performance} and Fig. \ref{fig:ine}, respectively.

\begin{figure}[t]
        \begin{center}
                \includegraphics[width=0.9\linewidth]{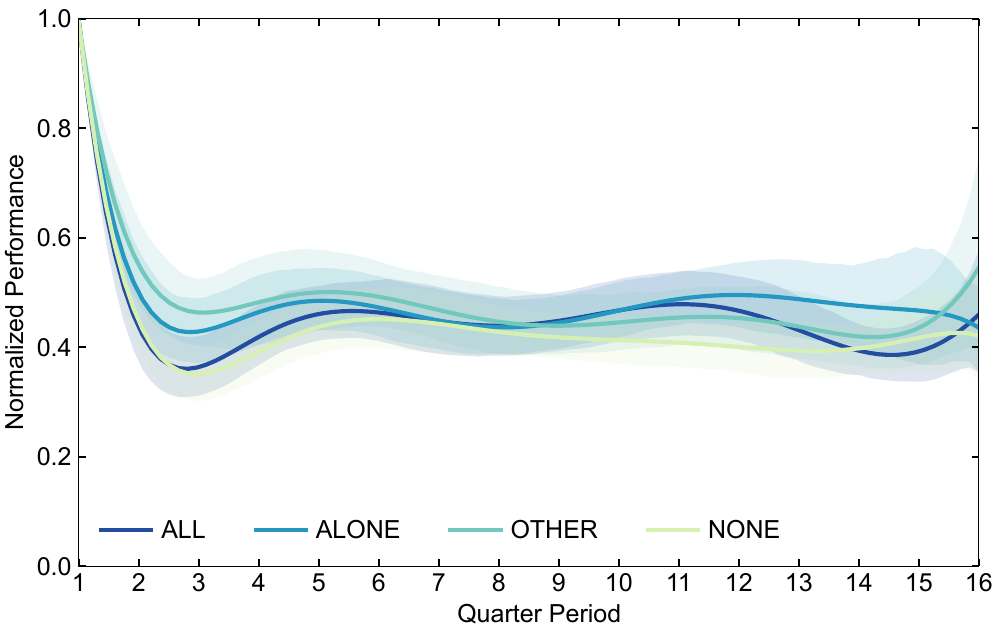}
                \caption{Comparison of the temporal changes in normalized performance.}
                \label{fig:normalized}
        \end{center}
\end{figure}

Additionally, changes in performance were observed over time.
Fig. \ref{fig:normalized} shows a comparison of the temporal changes in performance.
Here, a ``Quarter Period'' represents the average of each quarter of a rose curve, with the period of the rose curve being 20 s (\textit{i.e.} the one-quarter period is 5 s).
The plate control task was conducted for 80 s, thus the graph displays 16 quarter periods.
Performance for each quarter period was normalized based on performance in the first quarter period.

Consequently, it was confirmed that the dyads improved their performance compared to the first quarter period under ALL, ALONE, OTHER, and NONE conditions (lower values indicate better performance).
However, no clear differences were detected between the conditions ALL, ALONE, OTHER, and NONE.
Therefore, while performance improved during the collaborative task, the specific condition did not significantly influence the rate or extent of improvement.

\section{DISCUSSION}
\subsection{Subjective measures}
The survey revealed significant main effect in all measures of Social Presence, Co-presence, Perceived Attentional Engagement, and Perceived Behavioral Interdependence. 
Notably, significant differences were found in comparisons of the co-presence sub-scale, except between the ALONE and NONE conditions.
These results caused the rejection of the hypothesis ${\bf H1}$.

The ALL condition yielded the highest scores, indicating a high level of social presence, which was not unexpected.
It is well documented that the presence of a partner's avatar increases social presence, as demonstrated through various studies \cite{Aseeri2021-xw, Yoon2019-gi, Heidicker2017-wk, Weidner2023-ee}.
Our findings reinforce the importance of displaying a partner's avatar in the SVE.

Social presence scores in the OTHER condition were the next highest after ALL, but ALONE and NONE showed lower scores compared to ALL and OTHER.
Therefore, having a self-avatar is less critical for maintaining a high social presence than seeing the partner's avatar.
Participants were constantly engaged in collaborative tasks that involved plate handling during the experiments, a constant factor in all avatar conditions.
The partner's presence was sensed through haptic interaction regardless of the avatar's visibility.
However, the absence of the partner's avatar reduced the social presence.
Haptic feedback plays a vital role in SVEs, and several studies have reported that it improves social presence \cite{Sallnas2010-tp, Chellali2011-tf, Beelen2013-be, Fermoselle2020-ta, Giannopoulos2008-bv}. 
These results indicate that visual representation of a partner, such as avatars, might be more critical than haptic interaction for enhancing social presence.

Subsequently, we focus on the perceived behavioral Interdependence.
Our results indicate that the ALL condition scored significantly higher than NONE.
Perceived Behavioral Interdependence measures the extent to which users' actions affect and are affected by their partners.
Thus, participants felt their influence on or influence from the partner was more significant when avatars were visible.
This finding is supported by studies, such as those by Heidicker \textit{et al.} \cite{Heidicker2017-wk} and Yoon \textit{et al.} \cite{Yoon2023-vp}.
In our study, the dyads were constantly provided with haptic feedback, and their influence on each other always existed between the dyads, regardless of avatar visibility.
However, avatars influenced participants, potentially feeling monitored by their partners (similar to how one might behave more cautiously when a camera is on during a web conference).
This observation suggests that it may be possible to choose to omit the display of avatars to reduce the mental burden on users.

\subsection{Objective measures}
The measurements of performance and effort revealed no significant differences based on avatar representation.
Despite visual limitations, the dyads continued to demonstrate the level of performance that they could achieve in the plate control task.
Thus, hypothesis ${\bf H2}$ was supported. 
Various studies have investigated the impact of avatar representation on task performance, yielding mixed results \cite{Gao2020-kc, Pan2017-dl, Yoon2023-vp, Wu2021-pq}.
Yoon \textit{et al.} explained that the importance of avatars varies between user-centered situations (where the main focus is on collaboration among users) and task-centered situations (where the main focus is on completing the task through collaboration) \cite{Yoon2023-vp}.
Our plate control task, which could be completed irrespective of avatar representation, falls into the category of task-centered situations, and the consistent performance across different avatar conditions supports their claim.

Moreover, two studies that examined the impact of visual interaction in remote collaboration tasks pointed out that showing partner manipulation information as cursors on the screen improved the performance of dyad work \cite{Chackochan2019-sg, Lokesh2023-qo}.
The tasks in their studies required estimating the partner's maneuvers, where visual information about the partner was crucial for task completion.
Our plate control task involved coordinating the manipulation of a rigid plate, and the information provided through haptic interaction may have been sufficient to estimate the partner's actions.
If the task had involved moving a table, where hand positioning is critical, avatar representation might significantly impact performance, especially when the partner's physical information plays a crucial role in task completion.

In terms of effort, hypothesis ${\bf H3}$ was rejected.
Dyads did not alter their effort despite changes in visual information.
For physically connected dyads, the forces between partners play a crucial role, especially in negotiating and integrating intentions \cite{Groten2013-gt, Roche2022-id}.
Groten \textit{et al.} demonstrated that haptic information provides a valuable channel for integrating intentions between dyads in co-performed haptic tasks.
Roche \textit{et al.} supported this, explaining that when dyad intentions differ, the forces exerted increase, and haptic information is utilized in negotiation.
These findings suggest that the presence of an avatar does not influence users' intentions in performing tasks.
This is because the avatars used in our study were generic and lacked gender, expressions, gaze direction, and physical cues that could significantly influence interactions.
However, studies have shown that avatar appearance, such as gender, can influence the degree of force perceived to be appropriate for instance, interactions with female avatars might result in gentler forces compared to male avatars \cite{Bailenson2008-kn}.
Therefore, using avatars with a higher level of detail could potentially affect haptic interactions.

\subsection{Generalization}
We developed an argument that integrated subjective and objective measures.
Although changes in social presence were observed across the four avatar conditions, performance and effort remained unchanged.

The first key finding is that for dyads physically connected by haptic interaction, avatar representation does not impact performance or the quality of haptic interactions.
This insight could significantly influence the development of physical remote collaboration applications.
Implementing tracking systems for such systems may not be essential.
Typically, developing VEs that display avatars requires cameras or sensor-based tracking systems, which are subject to constraints, such as location and lighting \cite{Waltemate2018-kv}.
Additionally, communication and rendering technologies are needed, which contribute to the cost of system construction in both hardware and software \cite{Weidner2023-ee}.
Thus, our findings could cause reductions in various costs associated with these systems.

The second key finding is that the enhancement of social presence provided by avatar representation cannot be substituted by haptic interaction alone.
This finding is particularly relevant for the entertainment sector, where user experience is crucial.
Presenting avatar representations in real time to users in an SVE is sufficient to enhance the experience of emotional closeness, similar to that of audio and video \cite{Bente2008-sk}.
Therefore, avatars play a starring role in social communication channels.
Our findings indicate that, on the contrary, haptic interactions play a supporting role.
Consequently, avatar representation might be more important than haptic interaction for enhancing user experience, emphasizing the significance of visual elements in the perception of social presence and interaction quality.

This distinction highlights the critical role avatars play in the user's emotional and psychological engagement within VEs, suggesting a strategic focus on visual representation to enhance social presence effectively, particularly in applications where the social dimension is predominant.

Our experiment intentionally employed a controlled and abstract task to isolate the effects of avatar representation and haptic interaction. 
This design restricted non-verbal or verbal communication, which may not reflect real-world scenarios where participants often rely on multiple communication cues. 
Although such an experimental setup allowed us to distinct between the contributions of haptic feedback and avatar representation, it also limits the direct applicability of our findings to more naturalistic tasks that combine social and performance demands simultaneously.

For instance, tasks that involve both a strong social component (\textit{e.g.}, negotiation, empathy, or team building) and a high emphasis on performance \cite{Khojasteh2021-rw,Bian_undated-dt} may benefit from a more nuanced integration of avatars and haptic feedback.
In such scenarios, relying on haptic information might not suffice for effective collaboration, while simultaneously, the need for accurate task performance could make the absence of avatars less optimal\cite{Yoon2023-vp}.
Future research should investigate how richer communication channels—including voice, facial expressions, and higher-fidelity avatars—interact with haptic feedback when both social presence and task performance are equally important.
Chellali \textit{et al.} reported that haptic feedback induces changes in verbal communication, leading to an increase in discussions related to force. 
Such shifts in conversation can, in turn, reciprocally affect how haptic information is shared, and they are likely to occur in highly contextual social scenarios. 
Meanwhile, Roth \textit{et al.} compared the time required to reach an agreement in a negotiation role-play—a verbally oriented task—between a real-world setting and a VE using abstract avatars\cite{Roth2016-se}.
They found no significant differences, indicating that even abstract avatars can effectively support communication channels.
Therefore, investigating interactions that align with the specific context of social tasks is essential for achieving a comfortable remote collaboration.

In summary, while our findings suggest that avatar representation does not affect task performance in a strictly task-centered scenario, and that haptic feedback alone cannot substitute for visual cues in enhancing social presence, these conclusions are based on a controlled context that may not fully capture the complexity of real-world interactions. 
Researchers and designers should carefully consider whether their target application places greater emphasis on social presence, task performance, or both, and evaluate if additional communication modalities are needed beyond the basic haptic and visual channels examined here.

\section{LIMITATION AND FUTURE WORK}
\subsection{Level of Detail on avatar representation}
The avatars used in this study were generic and lacked detailed physical information, which was intentional for a preliminary investigation.
However, the level of detail (LoD) in avatar representation could significantly impact user experience.

The Proteus effect explains how changes in self-avatar appearance can influence user behavior and cognition \cite{Yee2007-mi}.
Use of muscular avatars has been reported to lead to reduced perceived effort and altered actions \cite{Lin2021-og, Kocur2020-yf}.
Furthermore, the use of personalized avatars can improve body ownership and immersion in virtual spaces \cite{Waltemate2018-kv, Genay2022-ko}.

In SVEs, avatar expression has a significant impact on users.
A study by Kang \textit{et al.} reported that a low LoD in facial expressions of avatars can diminish social presence \cite{Kang2022-ee}.
Conversely, interacting with expressive avatars can enhance social presence \cite{Wu2021-pq, Combe2024-ef, Oh2016-be}.
Moreover, using a 3D-scanned avatar was shown to yield a stronger sense of virtual body ownership (VBO) compared to an abstract avatar, underscoring the importance of avatar realism in social interactions \cite{Latoschik2017-mz}. 

Therefore, varying the appearance and expressions of avatars could extend the discussions presented in this paper. 
Typically, cameras are used to apply facial expressions to avatars, but integrating them with an HMD is a challenge \cite{Rogers2022-vl}.
Therefore, facial tracking devices that can be mounted on HMDs have been developed, allowing expressions to be represented even while wearing an HMD \cite{Li2015-op, Numan2021-sy}.
Recently, Apple Inc. has developed a personalized avatar system, known as Persona, which can express facial details using their Apple Vision Pro\texttrademark \cite{noauthor_undated-ax}.

In future studies, we plan to use these advanced devices to expand our investigation and further explore how enhanced avatar realism and expressiveness impact user experience in SVEs.
This approach may reveal deeper insights into the behavioral and psychological effects of high-fidelity avatars, potentially guiding the design of more engaging and effective VR experiences.

\subsection{The impact on group size}
In this paper, participants were limited to dyads, but there is potential to expand the research to groups larger than three people.
Generally, as the group size increases, the dynamics of the group become more pronounced, and various psychological effects occur.
For instance, larger groups might observe an increase in free riders and a decrease in individual proactivity and performance.
This decrease can be explained by phenomena, such as the Ringelmann effect or the bystander effect \cite{Ringelmann1913-wi, Darley1968-zy}.

In our investigation, we found only one study that investigated the impact of changes in group size on social presence within an SVE \cite{Moustafa2018-kx}.
Their research involved tasks where 2--4 person groups used HMDs to communicate, examining social presence.
They found that smaller groups tend to engage more closely with each other, experiencing a higher social presence.
However, larger groups might feel a certain distance in their interactions with others, which can affect the sense of social presence.

Consider these findings, it is clear that increasing group size could have significant impacts, but actual effects would need to be investigated.
However, our developed system supports collaborative work for up to four people, allowing future studies to explore the effects of group size.
This exploration could significantly contribute to understanding how social presence is mediated by group dynamics in VEs, potentially influencing the design of VR systems and applications for larger groups.

\subsection{Collaboration task}
The plate control task developed and used in this paper was a task that involved coordinated manipulation of a rigid body.
This task required technical and precise adjustment of the force.
Future research using different types of haptic sharing tasks may develop the discussion further.

Maroger \textit{et al.} conducted research on the task of carrying a table in dyads.
Carrying a table is a routine and seemingly simple context, however, they showed that the strategies for carrying a table vary by dyad and are complex \cite{Maroger2022-yu}.
Sawers \textit{et al.} conducted research with an interesting task of dancing in dyads \cite{Sawers2017-yv}.
Their study demonstrated the potential for small interaction forces to convey movement goals.

Thus, the impact on user experience may vary depending on the type of haptic sharing task.
Future research is expected to expand the discussions of this paper through various haptic sharing tasks.

\section{CONCLUSION} 
In this study, we investigated how avatar representation affects social presence and collaborative tasks in a Shared Virtual Environment (SVE) with haptic interaction. 
Four conditions of avatar representation were prepared: both the participant's and the partner's avatars are displayed (ALL), only the participant's avatar is displayed (ALONE), only the partner's avatar is displayed (OTHER), and no avatars are displayed (NONE). 
In all conditions, we used a generic, gender-neutral avatar to eliminate variables, such as facial expressions or gender cues. 
The experiment involved a plate control task with haptic interaction, which was intentionally designed as an abstract and controlled scenario to ensure consistency among conditions. 
The analysis was performed through subjective measures of social presence and objective measures of performance and effort. 
Social presence scored higher in ALL and OTHER conditions and lower in the ALONE and NONE conditions. 
Therefore, although haptic interaction provided the presence of a partner, the presence of the partner's avatar was suggested to be important for enhancing social presence. 
Performance and effort remained consistent regardless of avatar representation, indicating that avatar representation does not affect haptic interaction under these controlled conditions.
Consequently, we present the following contributions:
1) Avatar representation does not impact performance or effort of haptic interaction; therefore, avatars can be omitted when the sole purpose is physical collaboration.
2) Visual interaction through avatars plays a leading role in social communication channels, but haptic interaction can only play a supporting role.

We hope that the findings of this paper, although derived from a generic avatar and a controlled plate control task, will aid in the development of physical remote collaboration through SVEs.

\section*{Acknowledgments}
This work was partially supported by; Research Institute for Science and Technology of Tokyo Denki University Grant Number Q23D-07 and Q24D-01 / Japan and JSPS KAKENHI Grant Number 22H01455.

\bibliography{ref}
\bibliographystyle{IEEEtran}



\vfill

\end{document}